\documentclass{PoS}

\usepackage{amsmath}

\usepackage{textcomp}

\newcommand{\I}{\ensuremath{\mathrm{i}}}

\title{Clover fermions in the adjoint representation and simulations of supersymmetric Yang-Mills theory}

\ShortTitle{Clover fermions in the adjoint representation}

\author{Sven Musberg, Gernot M\"unster, \speaker{Stefano Piemonte} \\
       Universit\"at M\"unster, Institut f\"ur Theoretische Physik,\\
     Wilhelm-Klemm-Str. 9, D-48149 M\"unster, Germany\\
        E-mail: \email{spiemonte@uni-muenster.de}}

\abstract{Clover improvement is the standard choice for lattice simulations of QCD, when the lattice artefacts coming from Wilson fermions have to be reduced. However, the clover improvement is not limited to QCD, but can be applied to a wider range of theories with fermions in higher representations of the gauge group SU(N), like the adjoint fermions required by supersymmetry or by technicolor theories. We present the calculation of the clover coefficient up to one loop order with standard perturbation theory for these models. Applications of clover fermions to supersymmetric Yang-Mills theory are also discussed.}

\FullConference{31st International Symposium on Lattice Field Theory LATTICE 2013\\
                 July 29 - August 3, 2013\\
                 Mainz, Germany}

\begin{document}

\section{Introduction}

During the last decades, fermions in higher representations of the gauge group have been introduced in several theories for describing the physics beyond the standard model. For example fermions in the adjoint representation are the basic constituents of the strongly-interacting technicolor fields and of the gauge sectors of supersymmetric models.

Many efforts have been spent recently to study the low energy behavior of these theories with numerical simulations on the lattice \cite{giedt,patella,berg2}. Most of the computer time is usually spent for extrapolating to the limit $a \rightarrow 0$, where the Lorentz invariance, broken by the lattice, is recovered. The standard Wilson formulation of the fermions leads to discretization errors of order $O(a)$ and therefore to a slow convergence towards the continuum limit. An improved version of Wilson fermions can be constructed following the Symanzik program \cite{Symanzik1,Symanzik2}. The strategy is to add the irrelevant operator $O_{CL}$ to the unimproved Lagrangian \cite{WohlertOrig},
\begin{eqnarray}
 L_i & = & L_u + c_{sw} O_{CL} \\
 O_{CL} & = & - \frac{a}{4} \bar{\lambda}(x) \sigma_{\mu\nu} F^{\mu\nu} \lambda(x) \label{clover}
\end{eqnarray}
and to tune the coefficient $c_{sw}$ to an appropriate value where the lattice discretization errors $O(a)$ disappear from on-shell quantities (like masses of mesons or particle scattering cross sections). The operator $O_{CL}$ is called clover term, because of the shape depicted by the plaquettes of the lattice version of $F^{\mu\nu}$. The Sheikholeslami-Wohlert coefficient $c_{sw}$ is a function of the gauge coupling $g$ and it can be expanded asymptotically as:
\begin{equation}
 c_{sw} = c_{sw}^{(0)} + c_{sw}^{(1)} g^2 + c_{sw}^{(2)} g^4 + \dots
\end{equation}

The value of $c_{sw}$ can be computed in standard perturbation theory. In case of QCD, the fermions are in the fundamental representation of the gauge group SU($N$) and $c_{sw}$ has been computed both in perturbation theory up to one loop \cite{Wohlert,luscherpert} and non-perturbatively within the Schr\"odinger functional formalism \cite{luschernonpert,effectclovqcd}. Similar numerical calculations have been done for the two-flavor adjoint QCD \cite{kari}, but so far no perturbative calculation of $c_{sw}$ has been available. In this article we summarize the calculation of the Sheikholeslami-Wohlert coefficient up to one loop for fermions in arbitrary representations of the gauge group SU($N$), as presented in Ref.~\cite{mpert}. A particular interest of our investigations is the special case of the $\mathcal{N} = 1$ supersymmetric Yang-Mills theory.

\section{Lattice fermions in representations of SU($N$)}

\begin{figure}[t]
\centering
\includegraphics[width=0.27\textwidth]{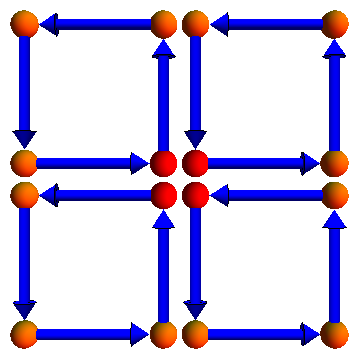}
\caption{The diagram represents the lattice version of $F_{\mu\nu}$ as the sum of the plaquettes based over the red point.}
\label{cloverimage}
\end{figure}

Let $R$ be an irreducible representation with dimension $d_R$ of the special unitary group SU($N$). Then the lattice link variables $U_\mu^R(x)$ are related to the continuum color gauge field $A_\mu^b(x)$ by the exponential map
\begin{equation}\label{expmap}
 U_\mu^R(x) = \exp{(\I g a A_\mu^b(x) \tau_b^R)},
\end{equation}
where $\tau_b^R$ are the Lie group generators in the representation $R$, obeying
\begin{equation}
 [\tau_b^R, \tau_c^R] = \I f_{bc}^d \tau_d^R\,.
\end{equation}

The generators $\tau_b^R$ are $N^2 - 1$ hermitian matrices of size $d_R \times d_R$. The fundamental representation has dimension $d_R = N$ and the matrices $\tau_d^F$ are normalized such that
\begin{equation}
 \textrm{Tr}(\tau_b^F \tau_c^F) = \delta_{bc}\frac{1}{2}\,.
\end{equation}

The generators in the adjoint representation $\tau^A_b$ are defined using the group structure constants as
\begin{equation}
 (\tau^A_j)_{kl} = -\I f_{jkl}\,.
\end{equation}

On the lattice, the gauge part of the action is written in terms of the links in the fundamental representation as
\begin{equation}
 S_g = \frac{\beta}{N} \sum_{x, \mu < \nu} \textrm{Re}\textrm{Tr}(U_\mu^F(x) U_\nu^F(x+\mu) U_\mu^F(x+\nu)^\dag U_\nu^F(x)^\dag)\,,
\end{equation}
while the fermion part depends on the links in the arbitrary representation $R$:
\begin{equation}
 S_f = \sum_{x,y} \bar{\lambda}(y) D_W[U_\mu^R](y,x) \lambda(x).
\end{equation}

The action of the Dirac-Wilson operator $D_W$ on the fermion field $\lambda$ is (Dirac and color indices suppressed)
\begin{equation}
 D_W(x,y)\lambda(y) = \lambda(x) - \kappa \sum_{\mu} \left\{ (1-\gamma_\mu) U_\mu^R(x) \lambda(x+\mu) + (1+\gamma_\mu) U_\mu^R(x-\mu)^\dag \lambda(x-\mu)\right\}.
\end{equation}

The clover term is defined in Eq.~(\ref{clover}) and $F_{\mu\nu}$ is chosen as the lattice antisymmetric sum of the plaquettes $P_{\mu\nu}$ of Fig.~\ref{cloverimage}
\begin{equation}
 F_{\mu\nu}(x) = \frac{1}{8\I} \sum_{\mu < \nu} (P_{\mu\nu}(x) - P_{\mu\nu}(x)^\dag).
\end{equation}

The Feynman rules can be obtained expanding the exponential map of Eq.~(\ref{expmap}), the complete list can be found in Ref.~\cite{aoki}.

\begin{figure}[t]
\centering
\includegraphics[width=0.39\textwidth]{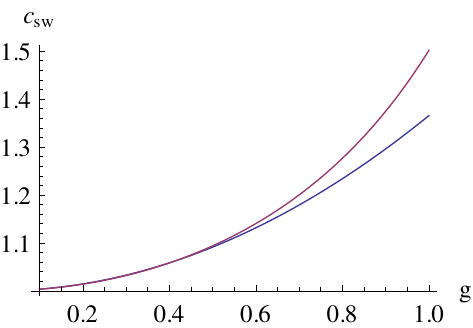}
\caption{The plot compares $c_{sw}$ between the one loop calculation (blue line) and the numerical estimation (purple line) for the two-flavor adjoint theory of Ref.~\cite{kari}.}
\label{clovercomp}
\end{figure}

\section{Calculation of the Sheikholeslami-Wohlert coefficient}

The calculation of the Sheikholeslami-Wohlert coefficient proceeds as in standard QCD. Many possible on-shell quantities can be considered for effectively performing the calculation of $c_{sw}$, the final computed value will be independent of this choice \cite{onshell}. In the original calculation of Wohlert the energy levels were considered, in our calculation we use instead the fermion-fermion scattering, following Ref.~\cite{aoki}.

The relevant part for the calculation of the clover coefficient comes from the fermion-fermion-gluon vertex $\Lambda(p,p')^\mu_c$
\begin{equation}
 \Lambda(p,p')^\mu_c = g \left(\I \gamma^\mu A + \frac{a}{2} (B - c_{sw}) + O(a^2) + O(p^2,p'^2)\right) \tau^R_c\,,
\end{equation}
where $p$ and $p'$ are the momenta of the incoming fermions and $c$ is the color of the outgoing gluon. The first term represents the continuum color vertex, while the second one is a lattice artifact proportional to $a$. At tree level one finds
\begin{equation}
 B = 1 + O(g^2)
\end{equation}
independent of the number of colors and of the fermion representation. Thus the clover coefficient must be set to the universal value $c_{sw}^{(0)} = 1$.

At one loop, the calculation of the vertex involves the reduction of products of matrices $\tau^R_a$,
\begin{eqnarray}
\sum_a \tau^R_a \tau^R_a & = & C_R \mathbf{1}, \\
\sum_{bc} f_{abc}\tau^R_b \tau^R_c & = & \I \frac{N}{2} \tau^R_a\,, \\
\sum_{b} \tau^R_b \tau^R_a \tau^R_b & = & \left(C_R - \frac{N}{2} \right)\tau^R_a\,,
\end{eqnarray}
so that the final results depend only on the number of colors of the gauge group and on the quadratic Casimir invariant $C_R$. The latter can be obtained using group theoretical formulas, see \cite{patella,group}.

At one loop the $O(a)$ lattice artifact coming from the vertex is
\begin{equation}
B = 1 + g^2(0.16764(3)C_R + 0.01503(3)N) + O(g^4),
\end{equation}
which can be removed if $c^{(1)}_{sw}$ is set to
\begin{equation}
c^{(1)}_{sw} = 0.16764(3)C_R + 0.01503(3)N.
\end{equation}

Numerical estimations of the clover term have been made in Ref.~\cite{kari} in the special case of a theory with two fermions in the adjoint representation of SU(2). The expansion of their results yields
\begin{equation}
c_{sw}^{(mc)}(g) = 1 + 0.346806g^2 + O(g^2)
\end{equation}
in good agreement with our perturbative value of $c^{(1)}_{sw} = 0.36533(4)$ (Fig.~\ref{clovercomp}).

\section{Applications to $\mathcal{N}=1$ SUSY Yang-Mills theory}

The $\mathcal{N}=1$ supersymmetric Yang-Mills theory describes the interactions between gluons and their supersymmetric partners, called gluinos. The gluino is a spin-\textonehalf~ Majorana fermion in the adjoint representation. A Majorana fermion obeys the condition
\begin{equation}
 \bar{\lambda} = \lambda^T C,
\end{equation}
where $C$ is the charge conjugation operator. As a consequence there is not a distinction between particle and antiparticle, i.\,e.\ the two-point Green functions $\langle \bar{\lambda}(x) \lambda(y) \rangle$, $\langle\bar{\lambda}(x) \bar{\lambda}(y) \rangle$ and $\langle \lambda(x)\lambda(y) \rangle$ do not vanish. The main difference appears in the combinatorial factors for each Feynman diagram that has fermion loops. The one-loop diagrams for the fermion-fermion-gluon vertex have no fermion loops and therefore the previous value of $c_{sw}^{(1)}$ can be applied also to Majorana fermions. The difference between Dirac and Majorana fermions is expected to be found at order $O(g^4)$.

Recent numerical investigations of $\mathcal{N}=1$ Super Yang-Mills theory have been presented in Refs.~\cite{berg1,berg2}, where the mass gap for particles in the same supermultiplet is found to be significantly reduced on finer and bigger lattices. The use of clover improvement could be helpful for simulating the model also on coarser lattices, since the finite lattice spacing $a$ is an important source of supersymmetry breaking. In fact, the SUSY algebra mixes the internal field symmetry with Poincar\'e transformations,
\begin{equation}
\{ Q_\alpha, Q_\beta\} = (\gamma^\mu C)_{\alpha \beta} P_\mu,
\end{equation}
where $P_\mu$ is the translation operator. On the lattice only discrete translations of step $a$ are possible, therefore the supersymmetry algebra is explicitly broken for a finite lattice spacing and it could be restored only in the continuum limit (considering only local actions, see \cite{curci}). The faster approach to the continuum limit given by the clover term can be helpful for restoring SUSY also on coarser lattices.

In order to be competitive with other possible choices, the addition of the clover term should not have a big impact on the performance of the code in the production of gauge configurations. The bottleneck of large scale supercomputer simulations is mostly related to the MPI communications between processors, from this point of view the clover term is a local operator and it does not require any other additional information to be shared. So the clover term can be used for achieving real improvement on the final results.

\section{Conclusions}

In this contribution the calculation of the improvement coefficient $c_{sw}$ to one-loop order is presented for supersymmetric Yang-Mills theory and models with Dirac fermions in any representation of the gauge group SU($N$). Further numerical simulations will be done for understanding the efficiency of the clover improvement for restoring supersymmetry on the lattice.

\end{document}